\pdfoutput=1 
\documentclass[english,11pt,a4paper]{article}

\usepackage{jheppub}
\usepackage{graphicx}
\usepackage{bm}
\usepackage{amsmath}
\usepackage{lineno}
\usepackage{slashed}
\usepackage{amssymb}
\usepackage[utf8]{inputenc}
\usepackage{lipsum}
\usepackage{stmaryrd}

\IfFileExists{dsfont.sty}
	{\usepackage{dsfont}
         \let\mathbb=\mathds
         \newcommand{\id}{\mathds{1}}}
	{\typeout{Package dsfont.sty was not found, using alternative macros.}
         \let\mathds=\mathbb
         \newcommand{\id}{\mbox{1 \kern-.59em {\rm l}}}}

\renewcommand\a{\alpha}
\renewcommand\b{\beta}
\renewcommand\d{\delta}

\renewcommand\l{\lambda}
\renewcommand\r{\rho}

\renewcommand\o{\omega}

\newcommand\e{\epsilon}
\newcommand\g{\gamma}

\newcommand\m{\mu}
\newcommand\n{\nu}
\newcommand\x{\xi}
\newcommand\p{\pi}

\newcommand\s{\sigma}

\renewcommand\S{\Sigma}
\renewcommand\O{\Omega}

\newcommand\G{\Gamma}

\newcommand{\eq}[1]{Eq.~(\ref{#1})}

\newcommand\lb{\left(}
\newcommand\rb{\right)}
\newcommand\ls{\left[}
\newcommand\rs{\right]}

\newcommand{\lag}{\langle}
\newcommand{\rag}{\rangle}

\newcommand\dd{\partial}
\newcommand{\na}{\nabla}

\newcommand{\cZ}{{\cal Z}}
\newcommand{\cF}{\mathcal{F}}
\newcommand{\cP}{\mathcal{P}}

\newcommand{\tr}{{\rm tr}}

\newcommand{\bx}{{\bm x}}

\newcommand{\bp}{{\bm p}}

\newcommand{\bq}{{\bm q}}
\newcommand{\bpsi}{\Bar{\psi}}

\newcommand{\ba}{{\bm a}}

\newcommand{\bb}{{\bm b}}
\newcommand{\bbe}{{\bm \b}}

\renewcommand{\part}{{\rm part}}

\renewcommand{\bq}{\begin{equation}}
\renewcommand{\eq}{\end{equation}}
\newcommand{\bear}{\begin{array}{cc}}
\newcommand{\ear}{\end{array}}
\newcommand{\ds}{\displaystyle}

\newcommand{\lrd}{\overset{\leftrightarrow}{\partial}}
\newcommand{\lrn}{\overset{\leftrightarrow}{\na}}

\newcommand{\half}{\frac{1}{2}}
\newcommand{\Sn}{\S_n}

\allowdisplaybreaks
\begin{document}

\title{Zilch Vortical Effect for Fermions}

\author[a,b]{Artem Alexandrov,}
\author[a]{Pavel Mitkin}

\affiliation[a]{ITEP, B. Cheremushkinskaya 25, Moscow 117218, Russia}
\affiliation[b]{Moscow Institute of Physics and Technology, Dolgoprudny, Moscow 141700, Russia}

\emailAdd{aleksandrov.aa@phystech.edu,pavel.mitkin@phystech.edu}

\abstract{We consider the notion of zilch current that was recently discussed in the literature as an alternative helicity measure for photons. Developing this idea, we suggest the generalization of the zilch for the systems of fermions. We start with the definition of the photonic zilch current in chiral kinetic theory framework and work out field-theoretical definition of the fermionic zilch using the Wigner function formalism. This object has similar properties to the photonic zilch and is conserved in the non-interacting theory. We also show that, in full analogy with a case of photons, the fermionic zilch acquires a non-trivial contribution due to the medium rotation - zilch vortical effect (ZVE) for fermions. Combined with a previously studied ZVE for photons, these results form a wider set of chiral effects parameterized by the spin of the particles and the spin of the current. We briefly discuss the origin of the ZVE, its possible relation to the anomalies in the underlying microscopic theory and possible application for studying the spin polarization in chiral media.}

\date{\today}
\maketitle

\section{Introduction}

It is well established by now that systems of chiral fermions exhibit a set of transport phenomena -- chiral effects -- which are closely tied with the axial anomaly of the underlying quantum field theory (QFT). These effects can considerably influence dynamics of a variety of systems from quark-gluon plasma to Weyl and Dirac semi-metals, see \cite{Huang:2015oca, Kharzeev:2015znc} for a review. The most known of these effects are the chiral magnetic effect (CME),  chiral separation effect (CSE) and chiral vortical effects (CVE) \cite{Kharzeev:2007jp,Fukushima:2008xe,Vilenkin:1979ui,Erdmenger:2008rm,Ambrus:2019khr}. In the case of Dirac fermions these are captured by the following equations:
\bq\bear
\label{effects}
\ds{ J^\a =  \frac{\m_5}{2\p^2} B^\a + \frac{\m\m_5}{\p^2} \O^\a} \\ \\
\ds{J^\a_5=  \frac{\m}{2\p^2} B^\a + \lb \frac{\mu^2+ \m^2_5}{2\pi^2}+\frac{T^2}{6}\rb\Omega^\a}
\ear \eq 
where $\mu,\,\m_5$ are the vector and axial chemical potentials correspondingly, $\Omega^\mu=\frac{1}{2}\epsilon^{\mu\nu\alpha\beta}u_\nu\partial_\alpha u_\beta$ is the vorticity vector, $B^\m = \half \e^{\m\n\a\b} u_\n F_{\a\b}$ is a magnetic field in the rest frame of the fluid element, and $u_\mu$ is the 4-velocity of the fluid. 

The origin of different parts in (\ref{effects}) has been widely discussed in the literature. The anomalous nature of magnetic effects can be demonstrated in a number of ways \cite{Landsteiner:2016led}. The connection of CVE to the anomaly is less evident. Indeed, the vortical effects survive even in the absence of electromagnetic field, when the expectation value of the anomalous divergence is zero. Nevertheless, the connection of the ``$\m^2$-terms'' to the axial anomaly is established in the literature, see for example \cite{Son:2009tf,Sadofyev:2010is}. 

The discussion on the origin of $T^2$-term in (\ref{effects}), which is sometimes called thermal CVE (tCVE), however, is still going on \cite{Landsteiner:2011cp,   Golkar:2012kb, Avkhadiev:2017fxj,Glorioso:2017lcn, Flachi:2017vlp,Landsteiner:2011iq,Stone:2018zel,Prokhorov:2020okl}. It is worth to mention that tCVE is present even for the fermions of zero charge. For instance, the approaches developed in \cite{Son:2009tf,Sadofyev:2010is} do not capture this part of CVE\footnote{In \cite{Son:2009tf} the corresponding part of the vortical conductivity can appear as a free parameter, not fixed by the anomaly}. However, there is a conjecture that the origin of tCVE is rooted in the mixed gravitational anomaly  \cite{Landsteiner:2011cp, Landsteiner:2011iq,Stone:2018zel}. This point of view is opposed by the general statement that tCVE as an effect of polarization of massless particles and, as it is present in the thermal equilibrium without any interaction, it seems to have no connection to anomalies \cite{Buzzegoli:2017cqy}. The possible way to relate these effects of polarization to the anomalies is to say that the state of thermal equilibrium is a result of the evolution of the system governed by the interactions that are anomalous, see e.g. \cite{Stone:2018zel}. 

If there is any relation between the CVE and gravitational anomalies, then, since similar anomalies are present for the fields of higher spins \cite{Dolgov:1988qx,Christensen:1978md}, there should be analogous effects for any fields with nonzero spin. Indeed, it is known that the CVE, understood as separation of particles of different helicities in response to a rotation, is present for photons, for which the local measure of helicity is given by the Chern current $K^\m =  \e^{\m\n\r\s} A_\n \dd_\r A_\s$ \cite{Avkhadiev:2017fxj,Prokhorov:2020npf,Prokhorov:2020okl}. Later this result was reproduced in the framework of chiral kinetic theory (CKT) \cite{Yamamoto:2017uul} and then was generalized for the case of particles of an arbitrary spin \cite{Huang:2018aly}.

The problem one has to overcome when discussing the photonic CVE is that current $K^\m = \e^{\m\n\r\s} A_\n \dd_\r A_\s$ is not a gauge-invariant object. One way to overcome this difficulty was recently proposed in \cite{Chernodub:2018era}. The idea is to use a different measure of helicity, which is known as the zilch currents \cite{lipkin, kibble}. These currents in their covariant form are actually tensors of an odd rang $s$, which we would refer to as a spin of the current, and the corresponding charge measures the difference between the number of photons of different helicity weighed with the energy of this photons in power $s-1$. As it was shown in \cite{Chernodub:2018era}, these currents do obtain a nonzero expectation value in response to a rotation of the system -- zilch vortical effect (ZVE). In \cite{Huang:2020kik} these zilch currents was considered in CKT and it was shown that ZVE, just like the usual photonic CVE, is connected to a non-trivial topological Berry phase.

It is tempting to generalize the notion of the zilch currents to the case of systems of particles of a different spin. In this paper we do that for the case of fermions and introduce a fermionic zilch -- a family of currents parameterized by an odd integer $s$ and conserved in the theory with no interaction. We start with a CKT definition of the photonic zilch from \cite{Huang:2020kik} and generalize it trivially to the case of fermions. Then we use the Wigner function formalism to work out the field-theoretical definition of these currents and find

\bq \label{zilchdefinition}
\cZ_{\a_1..\a_s} =  \bpsi \g_{\{\a_1} \lrd_{\a_2} .. \lrd_{\a_s\}} \g_5\psi \,.
\eq
where the index symmetrization is defined as $A_{\{\a_1...\a_s\}}=\frac{1}{s!}\sum A_{\Pi(\a_1...\a_s)}$ with the sum going over all permutations $\Pi$ and two-way derivative -- as $\lrd=\frac{1}{2}\big(\overset{\rightarrow}{\partial}-\overset{\leftarrow}{\partial}\big)$. In full analogy with the case of photons, fermionic zilches can serve as a measure of fermion chirality and have a non-trivial response to a rotation in thermal equilibrium -- fermionic ZVE, which depends both on the temperature of the system $T$ and the chemical potential $\m$, and is parameterized by the integer $s$. In the particular case $s=1$, when definition (\ref{zilchdefinition}) reduces to the usual axial current, this result reproduces the usual CVE (\ref{effects}). Finally, we use this field-theoretical expression and rederive ZVE via Kubo formula.

The notion of fermionic zilch and fermionic ZVE has not been studied in the literature. Since the origin of these novel effects in CKT is in complete analogy with the usual CVE origin, they could prove useful for studying the connection of CVE to quantum anomalies. In particular, if there is a relation between the usual CVE and gravitational anomaly, we expect that, first, there are gravitational anomalies in the zilch currents as well and, second, the relation between ZVE and these anomalies should be the same.


\section{Zilch currents in CKT}
\label{Sec:2}
We start with a general consideration of zilch currents in the framework of kinetic theory for massless particles. The comprehensive exposition of the fermionic case of CKT can be found in \cite{Chen:2014cla,Chen:2015gta}. This construction can be extended to the general case of a massless particle with an arbitrary spin \cite{Huang:2018aly}. We will review briefly some relevant points concerning the Lorentz-invariance of this theory. In the semi-classical limit the massless particle of spin $\half$ can be described by the phase-space action
\bq 
\mathcal{I} = \int (\bp \cdot d\bx - \e_\bp dt - \ba_\bp \cdot d\bp)\,,
\eq
where $\e_p$ is the energy of the particle and  $\ba_\bp$ is a Berry connection such that 
\bq
\bb = \bm \na_\bp \times \ba_\bp = \pm \frac{\bp}{2|\bp|^3}\,.
\eq 
It can be shown \cite{Chen:2014cla} that this action is not invariant under the ``naive'' Lorentz symmetry and in order to achieve it up to the first order in semiclassical expansion one has to modify the action of the boost generators in the following way:
\bq \label{sidejump}
\d_\bbe \bx = \bbe t + \half \frac{\bbe\times\bp}{|\bp|^2}\,; \quad \quad \d_\bbe t = \bbe \cdot \bx\,; \quad \quad \d_\bbe \bp = \e_\bp \bbe\,.
\eq 
The new contribution in the $\d_\bbe \bx$ is the infinitesimal version of the so-called side-jump \cite{Skagerstam:1992er, Chen:2014cla, Duval:2014ppa, Chen:2015gta}. This phenomenon is related to the ambiguity in the definition of the spin tensor of a massless particle. If we consider the total angular momentum tensor
\bq
J_{\m\n} = x_\m p_\n - x_\n p_\n +\S_{\m\n}\,,
\eq 
we can see that change in the definition of $\S_{\m\n}$ should be compensated by the corresponding shifts in $x^\m$ in order for $J_{\m\n}$ to be the same. It can be shown \cite{Chen:2015gta} that if we choose the frame by introducing some time-like unit vector $n^\m$ and fix the ambiguity in $S_{\m\n}$ by imposing an additional condition  $n^\m \S_{\m\n} = n^\n \S_{\m\n} = 0$, then the shifts in $x^\m$ induced by the change of the frame (change of the vector $n^\m$) is exactly the finite version of (\ref{sidejump}). For a chosen frame-vector $n^\m$ the spin tensor is fixed as
\bq
\Sn^{\m\n}=\l\,\,\frac{\e^{\m\n\r\s}p_\r n_\s}{p\cdot n}\,,
\eq
where $\l$ is a particle helicity.

Due to the shift in $x^\m$ the particle distribution function $f(x,\,p)$ is not a Lorentz scalar and is also frame dependent. This in turn leads to a modification in the definition of the phase space current density $j^\m (x,\,p)$ \cite{Chen:2015gta}:
\bq
j^\m=p^\m f+\Sn^{\m\n}\dd_\n f\,,
\label{CKTj}
\eq
In equilibrium, the distribution function $f(\x)$ depends on a linear combination of the integrals of motion $\x=\b_\m p^\m - \b \m +\frac{1}{2}\Sn^{\m\n}\O_{\m\n}$, where $\b^\m$ is the temperature 4-vector satisfying $T^2\b^\m\b_\m=1$, $\m$ is a chemical potential\footnote{Here we consider only the vector chemical potential which does not depend on polarization $\l$}, and $\O^{\m\n}=\frac{1}{2}\left(\dd^\m\b^\n-\dd^\n\b^\m\right)$ is the thermal vorticity tensor, see e.g. \cite{Chen:2015gta}. The expression for the current (\ref{CKTj}) is then frame independent and in first order in the semiclassical expansion reads:
\bq 
j^\m=p^\m f(\b_\n p^\n)-\half \l\, \e^{\m\n\r\s}p_\n\O_{\r\s} f'(\b\cdot p - \b \m)\,.
\eq
Now we can turn to the consideration of zilch currents in this framework. We start with a CKT definition of the zilch for photons. It should be noted that the tensor structure of the photonic zilch can be defined in different ways \cite{Copetti:2018mxw}. In CKT it is convenient to work with a fully symmetrical version. The expression for phase density for this version of the photonic zilch current of spin $s$ in terms of (\ref{CKTj}) was deduced in \cite{Huang:2020kik} and reads
\bq 
z^{(s,\l)}_{\a_1..\a_s}=(-1)^{\frac{s}{2}-\l}p_{\{\a_1}p_{\a_2}...j_{\a_s\}}\,.
\label{CKTzilchSP}
\eq
This expression can be generalized on systems of particles of arbitrary spin as is. The only difference in the fermionic case in comparison with the photons (apart from $\l$) is the distribution function $f$. Leaving the question of how to define the fermionic zilch current in QFT for the following section we write its expectation value in the CKT in a rotating system of fermions:
\bq
\label{general zve}
      \lag:\!\cZ^{i0 .. 0}(x=0)\!:\rag =  
    2(-1)^{\frac{s+1}{2}} \frac{(s+2)}{3s} \frac{\O^i}{T}\int \frac{d^4p}{(2\pi)^3}\d(p^2) p_0^{s} f'(\b\cdot p - \b \m)
\eq 
The explicit answers for the simplest cases of $\m = 0$ and $T=0$ read: 
\bq \begin{gathered}
\lag:\!\cZ^{i0 .. 0}(x=0)\!:\rag =   \frac{(-1)^{\frac{s-1}{2}}}{3\p^2} \O^i T^{s+1}\lb 1 - \frac{1}{2^{s}} \rb \frac{(s+2)(s+1)!}{s} \zeta(s+1)\,,
\end{gathered} \eq 
and 
\bq \begin{gathered} 
    \lag:\!\cZ^{i0 .. 0}(x=0)\!:\rag =    \frac{(-1)^{\frac{s-1}{2}}}{6\p^2}\frac{(s+2)}{s}\O^i \m^{s+1}\,.
\end{gathered} \eq
 Note that these results reproduce the (\ref{effects}) in case $s=1$.

\section{Wigner function consideration}

\label{Sec:3}

The general feature of calculations in CKT is the challenge of defining the phase space density corresponding to a given observable in the field theory. One way to deal with it is to use some formalism that connects CKT with the underlying field theory as it was done in \cite{Huang:2020kik} for the photonic zilch using the Wigner function formalism. Here we reverse this problem and recover the field-theoretical definition of the fermionic zilch from the definition in the kinetic theory (\ref{CKTzilchSP}). For that we use Wigner function for fermions that has been extensively explored in the literature \cite{Chen:2012ca, Gao:2012ix,Hidaka:2016yjf,Huang:2018wdl,Gao:2018wmr,Gao:2019znl,Prokhorov:2018qhq,Liu:2018xip, Hattori:2019ahi,Wang:2019moi,Weickgenannt:2019dks,Liu:2020flb}. Following these works, we define Wigner function for massless fermions as
\bq
{W}_{ab}(x,p) = \int \frac{d^4 y}{(2\pi)^4} e^{-ipy} \lag:\!\bpsi_b\lb x+\half y\rb  \psi_a \lb x-\half y\rb\!:\rag \,.
\eq
If interactions are negligible, this function should satisfy the following equations of motion:
\bq 
\label{equations}
\begin{cases}   \g^\m\lb p_\m + \frac{i}{2}\dd_\m\rb  W(x,p) = 0 \\ 
                W(x,p) \g^\m \lb p_\m - \frac{i}{2}\overleftarrow{\dd}_\m\rb  = 0.
\end{cases}
\eq
It is convenient to decompose $W_{ab}$ in terms of the generators of Clifford algebra
\bq 
W(x,p) = \frac{1}{4} \lb \cF + i\g_5\cP +\g^\m V_\m + \g_5 \g^\m A_\m + \half \s^{\m\n} S_{\m\n} \rb,
\eq 
where $\s^{\m\n} = \frac{i}{2}\ls\g^\m,\g^\n \rs $, and $\g_5 = -\frac{i}{4!} \e_{\m\n\a\b} \g^\m\g^\n\g^\a\g^\b$. In terms of these functions the equations (\ref{equations}) read:
\bq \bear 
        &  p^\m \cF - \half \dd_\n S^{\n\m} = 0; \quad \quad p^\m \cP + \frac{1}{4}\e^{\m\n\a\b}\dd_\n S_{\a\b} = 0; \\ \\
        &  \dd_\m \cF + 2 p^\n S_{\n\m} = 0; \quad \quad \dd_\m \cP - \e_{\m\n\a\b}p^\n S^{\a\b} = 0; \\ \\ 
        &  p_\m R^\m = p_\m L^\m = \dd_\m R^\m = \dd_\m L^\m = 0; \\ \\
        &  \half \dd_{[\m }R_{\n]} - \e_{\m\n\a\b} p^\a R^\b = 0; \\ \\ 
        &  \half \dd_{[\m }L_{\n]} + \e_{\m\n\a\b} p^\a L^\b = 0;
\ear \eq 
where index anti-symmetrization is defined as $a^{[\m} b^{\n]}=a^\m b^\n-a^\n b^\m$, and we introduced functions $R^\m = \half \lb V^\m + A^\m \rb$ and $ L^\m = \half \lb V^\m - A^\m \rb $, which represent the phase space currents of the right- and left-handed particles correspondingly. The solution up to the first order in semiclassical expansion can be found, for instance, in \cite{Gao:2012ix, Huang:2018wdl,Liu:2018xip} and we just use it:
\bq
\label{Rm}
R^\m = \frac{1}{(2\p)^3} 2\d(p^2) \lb p^\m +  \Sn^{\m\n} \dd_\n \rb f \,,
\eq 
where $\Sn^{\m\n}$ is the spin tensor with $\l = +\half$. The function $f(\x)$ satisfies Boltzmann equation $\d(p^2)\,p^\m\dd_\m f = 0$ and $\x=\b_\m p^\m - \b \m +\frac{1}{2}\Sn^{\m\n}\O_{\m\n}$. The solution for $L^\m$ is the same up to a minus sign in $\l$.  Using these solutions we can now easily work out the QFT form of the zilch current. Indeed, looking at the equation (\ref{CKTzilchSP}) we can write
\bq \begin{gathered}
\lag:\!\cZ^{\a_1 .. \a_s}(x)\!:\rag =   (-1)^{\frac{s-1}{2}}\int \frac{d^4 p}{(2\p)^4}  p^{\{\a_1} ..  p^{\a_s-1} \lb R^{\a_s\}}(x,p) - L^{\a_s\}}(x,p) \rb \\ =(-1)^{\frac{s-1}{2}} \int \frac{d^4 p}{(2\p)^4}  p^{\{\a_1} ..  p^{\a_s-1}  \tr \g^{\a_s\}} \g_5 W(x,p) \,,
\end{gathered}\eq 
where $\tr$ denotes the trace over spinor indices. The last line fixes zilch to be
\bq 
\label{fermizilch}
\cZ_{\a_1..\a_s} =\bpsi \g_{\{\a_1} \lrd_{\a_2} .. \lrd_{\a_s\}} \g_5\psi \,.
\eq
We see that in the case of $s=1$ this definition directly translates to the usual definition of axial current.

\section{Kubo formula approach}
\label{Sec:4}

Having the definition of zilch in terms of the fundamental fields we can perform an additional check for the equation (\ref{general zve}) using Kubo formula for linear response to the rotation \cite{Landsteiner:2013aba}. We define the vortical conductivity as
\bq 
\label{sigma def}
\lag:\!\cZ^{i0..0} \!:\rag= \s \O^i \, .
\eq
Considering the rotation as an external gravitational field, we can calculate this conductivity as a sum of two parts: a dynamical response that is defined by retarded correlator of $\cZ_{i0..0}$ with the stress-energy tensor of the system and a contact term which is contributed by the inner dependence of zilch definition (\ref{fermizilch}) on the metric, $\s = \s_T + \s_C$. The expressions for them are: 
\bq \begin{gathered}
\label{Kubo} 
    \lag \cZ^{i0..0} T^{0j} \rag_R|_{\o=0} = \frac{i}{2} \e^{ijk} p^k \s_T + O(p^2) \\
    \Big\lag \dfrac{\d \cZ^{i0..0}}{\d g^{oj}}\Big\rag  = \frac{i}{2} \e^{ijk} p^k \s_C + O(p^2) \,,
\end{gathered}\eq
where
\bq 
T_{0i} = \frac{i}{2} \bpsi\lb \g_0 \lrd_i + \g_i \lrd_0 \rb \psi 
\eq
In practice the contact term can be calculated directly as a one-point function if we know how the metric enters the definition of the zilch. In the external gravitational field we have to make (\ref{fermizilch}) covariant by changing the derivatives:
\bq 
 \cZ_{\a_1..\a_s} = \bpsi \g_{\{\a_1} \lrn_{\a_2} .. \lrn_{\a_s\}} \g_5\psi \,.
\eq
There are two sources for the metric in this equation: the first -- when the connection $\na$ acts on the Lorentz indices and the second comes from acting on the fermion field. It is easy to see, however, that when calculating the one-point function only the second source gives a nonzero contribution after computing the trace of gamma matrices. The spin connection for the fermion field can be written locally as $\na_\m = \dd_\m + \G_\m$ and in case of rotating frame it can be chosen to have only one non-zero component: $\G_0 = -\frac{i}{2}\O \s^{12}$ (we imply that the system rotates along $z$-axis). 

Without going into much detail we provide the expressions for both parts of the conductivity in terms of the moments of the distribution function: 
\bq\begin{gathered}
\label{moments kubo}
    \s_T =  \dfrac{(-1)^{\frac{s-1}{2}}}{2\pi^2} \frac{(s+1)}{s} \int\limits_0^\infty dq\, q^{s} \lb n_F(q-\m)+ n_F(q+\m) \rb\,; \\
    \s_C =  \frac{(-1)^{\frac{s-1}{2}}}{2\pi^2}  \frac{(s-1)(s+1)}{3s} \int\limits_0^\infty dq\, q^{s} \lb n_F(q-\m)+ n_F(q+\m) \rb\,,
\end{gathered}\eq
where $n_F(q) = \frac{1}{1+e^{\b q}}$ is a Fermi distribution function. Combining both contributions we have 
\bq 
\label{answer Kubo}
    \s = \dfrac{(-1)^{\frac{s-1}{2}}}{2\pi^2} \dfrac{(s+1)(s+2)}{3s} \int\limits_0^\infty dq\, q^{s} \lb n_F(q-\m)+ n_F(q+\m) \rb\,,
\eq
which coincides with the result obtained in kinetic theory. Using Jonquiere inversion formulae for the polylogarithm functions, we can write the general answer:

\bq 
\s = -\frac{(2\p)^{s+1}}{2\p^2} \frac{(s+2)}{3s} T^{s+1} B_{s+1} \lb \half + \frac{\m}{2 i \p T} \rb \,,
\eq 
where $B_k(x)$ are the Bernoulli polynomials.
\section{Conclusion}

In this work we have generalized the notion of the zilch current for the fermionic systems and showed that it obtains a non-trivial contribution in the case of a rotating fermion gas -- the fermionic ZVE. 

In Section \ref{Sec:2} we use the results from \cite{Huang:2020kik} and the generality of the kinetic theory approach to construct the phase density for the fermionic zilch and calculate its response to the rotation (\ref{general zve}). Naturally, just as it was shown for the photons in \cite{Huang:2020kik}, obtained effect (ZVE) can be traced back to the topological Berry phase and therefore has the same origin as the usual fermionic CVE, which is a particular case of ZVE. Combined with the previous results \cite{Chernodub:2018era,Huang:2020kik}, this gives us a whole new family of chiral effects parameterized by the spin of the zilch current and by the constituents spin, which can prove useful in studying the nature of the chiral effects.

In Section \ref{Sec:3} we use the Wigner function formalism to work out the field-theoretical definition of the fermionic zilch from the CKT (\ref{fermizilch}). Then in Section \ref{Sec:4} we use this to perform an independent check of the kinetic theory results, calculating the same effect via the Kubo formula. The results do coincide, thus supporting the CKT consideration and the field-theoretical definition of the zilch.

It should be stressed that the family of the zilch currents given by (\ref{fermizilch}) conserves only in the absence of interactions. It could be interesting to consider the dynamics of these quantities in the presence of, say, electromagnetic interaction and see if there is any interplay between the photonic and fermionic zilch currents. In particular, it is interesting to study how the non-conservation of the zilch turns into the axial anomaly in the limit $s=1$. The notion of the zilch can also be useful for studying the instabilities in chiral media, see e.g. \cite{Akamatsu:2013pjd, Khaidukov:2013sja, Kirilin:2013fqa, Avdoshkin:2014gpa, Manuel:2015zpa, Buividovich:2015jfa, Yamamoto:2015gzz, Hirono:2015rla, Kirilin:2017tdh, Li:2017jwv, Tuchin:2019gkg}. Another interesting observation is that one can also define a ``vector'' zilch, which would be defined by (\ref{zilchdefinition}) without $\g_5$. This current also conserves in the absence of interaction, and should obtain some response to a rotation if the axial chemical potential $\m_5$ is present.

There also is a question of the anomalies in the zilch currents themselves. In particular, it would be interesting to discuss the anomaly in the zilch arises in external gravitational fields. Returning to the discussion in the introduction we can say that if the tCVE is related to the mixed gravitational anomaly of the fermions, it can be argued that, first, there exists a similar anomaly in the fermionic zilch and, second, it is related in the same way to the ZVE. Seeing that ZVE is parameterized by the spin of the current $s$, this should prove useful in checking this conjecture; if there is a setup where this relation is invalid for some $s$, this would be an argument against the relation between the tCVE and the anomaly as well.

Finally, the study the zilch currents can be interesting in light of experimental measurements of hadron spin polarization in off-central heavy-ion collisions at RHIC and LHC~\cite{STAR:2017ckg,Adam:2018ivw,Acharya:2019vpe}.  The final state polarization follows the spin polarization of quarks and gluons in the QGP which, in turn, is in correspondence with the helicity transport due to the chiral effects. Introducing a new effect into the discussion can help in future studies of the spin polarization in heavy-ion collisions. 

\section{Acknowledgements}

The authors are thankful to A. Sadofyev, V.I. Zakharov, X.-G. Huang, K. Landsteiner and M. Chernodub for useful discussions. The work on this paper is supported by RFBR Grant 18-02-40056.  The work of P.M. is also supported by the Foundation for the Advancement of Theoretical Physics and Mathematics ``BASIS'' No.~20-1-5-134-1.

\appendix

\bibliographystyle{bibstyle}
\bibliography{zilch}
\end{document}